\documentclass{PoS}

\let\ifpdf\relax
\usepackage{ifpdf}

\def\bea{\begin{eqnarray}}
\def\eea{\end{eqnarray}}
\def\nn{\nonumber}
\def\msbar{$\overline{\rm MS\kern-0.5pt}\kern0.5pt$}
\def\lsim{\mathrel{\rlap{\lower4pt\hbox{\hskip1pt$\sim$}}\raise1pt\hbox{$<$}}}

\title{$SU(3)$ gauge theory with sextet fermions}

\ShortTitle{$SU(3)$ gauge theory with sextet fermions}

\author{\speaker{D\'aniel N\'ogr\'adi} \\
Institute of Physics, E\"otv\"os University, P\'azm\'any P\'eter s\'et\'any 1/a, Budapest 1117, Hungary \\
\email{nogradi@bodri.elte.hu}}

\abstract{
$SU(3)$ gauge theory coupled to $N_f = 2$ fermions in the sextet representation is a promising candidate for a technicolor inspired
Standard Model extension. In this note the progress in the past few years aimed at understanding the non-perturbative properties
of the model is reviewed. The main difficulties lying ahead in order to make robust conclusions from lattice simulations are
outlined.
}

\FullConference{The XXIX International Symposium on Lattice Field Theory - Lattice 2011\\
July 10-16, 2011\\
Squaw Valley, Lake Tahoe, California}

\begin{document}

\section{Introduction}
\label{introduction}

Lattice studies of technicolor inspired strongly interacting extensions of the Standard Model have seen considerable increase in
recent years. In particular $SU(3)$ gauge theory coupled to $N_f = 2$ flavors of sextet representation fermions was studied by
various methods and techniques. There are three main reasons for the interest in this particular model: perturbation theory
predicts that it lies close to the conformal window and may perhaps give rise to a walking coupling constant, the number of would-be Goldstone
bosons, if chiral symmetry is spontaneously broken, is exactly three and finally the number of flavors is low enough to hopefully give rise
to a low $S$-parameter. Taken together these expectations make $SU(3)$ gauge theory with $N_f = 2$ sextet fermions an appealing
candidate for strongly interacting extensions of the Standard Model. The role higher dimensional representations and the sextet in
particular may play in strong dynamics has been emphasized in \cite{Sannino:2004qp, Dietrich:2006cm}.

The 3-loop $\beta$-function of the model for $N_f$ flavors of Dirac fermions is
\bea
\mu \frac{dg}{d\mu} = \beta_1 \frac{g^3}{16\pi^2} + \beta_2 \frac{g^5}{(16\pi^2)^2} + \beta_3 \frac{g^7}{(16\pi^2)^3} \nn
\eea
\bea
\beta_1 = - 11 + \frac{10}{3} N_f\,,\qquad \beta_2 = - 102 + \frac{250}{3} N_f\,,\qquad \beta_3 = - \frac{2857}{2} +
\frac{30475}{18} N_f - \frac{11425}{54} N_f^2\,, \nn
\eea
where the first two coefficients \cite{Gross:1973id, Politzer:1973fx, Caswell:1974gg, Jones:1974mm, Banks:1981nn} 
are universal and the third  \cite{Tarasov:1980au, Larin:1993tp} is given in \msbar. Results on the 4-loop $\beta$-function
with higher dimensional representations can be found in \cite{Pica:2010xq}.
Asymptotic freedom forces $N_f < 3.3$. The $N_f = 1$ theory does not have spontaneous symmetry breaking because the
anomaly breaks the whole flavor group. This leaves the available interesting choices as $N_f = 2$ and $N_f = 3$ leaving aside half
integer $N_f$ values corresponding to Weyl fermions. 

The $N_f = 3$ model is close to the
upper end of the conformal window and is expected to be conformal. The fixed point coupling in the 2-loop
approximation is $g_*^2 / 4\pi \approx 0.085$ and it is expected that at such a small coupling the 2-loop result is trustworthy.
Indeed, taking into account the 3-loop correction the fixed point moves to $g_*^2 / 4\pi \approx 0.079$ which is only a $7\%$
change. 

The 2-loop result also predicts that the $N_f = 2$ model is conformal but the fixed point coupling $g_*^2 / 4\pi \approx 0.842$
is strong enough to doubt the validity of the 2-loop approximation. Again taking into account the 3-loop $\beta$-function the
fixed point moves to $g_*^2 / 4\pi \approx 0.5$ which is a $40\%$ change relative to the 2-loop result; it can hardly be
considered a small perturbation. In fact the ladder resummation in the 
Schwinger-Dyson approach predicts that the $N_f = 2$ model is actually chirally broken \cite{Sannino:2004qp, Ryttov:2010iz}. 

If chiral symmetry is broken for $N_f = 2$ the expected pattern of chiral symmetry breaking is $SU(2) \times SU(2) \to
SU(2)$ because the sextet representation is complex. This would result in three Goldstone bosons precisely the right number for
the three massive electro-weak gauge bosons.

In order for a model to be viable from a phenomenological point of view the $S$-parameter needs to be not too high otherwise it
would conflict with precision data \cite{Peskin:1990zt}. The phenomenon of walking requires the model to be just below the conformal window and
achieving this typically involves the increase of the fermionic content \cite{Appelquist:1986an, Appelquist:1986tr,
Appelquist:1987fc}. For the fundamental representation the interesting range
is then around $N_f = 10 - 14$ whereas for the sextet, as we have seen is around $N_f = 2$. A lower $N_f$ hopefully means a lower
$S$-parameter even with increased dimension for the representation, although this statement is also only valid in perturbation theory.

The above listed appealing features are all based on perturbation theory and/or analogy with QCD. Non-perturbative simulations are
then needed to clarify which of these expectations do hold. Before more detailed questions can be asked the basic property whether
in the infrared the model is chirally broken or conformal needs to be settled. Ideally, various approaches should yield the same
conclusion. The various approaches include different physics aspects (thermodynamics, mass spectrum, running coupling, etc) and
different regularization details (Wilson fermions, staggered fermions, improved vs. unimproved, etc). Agreement between different
regularizations is of course only expected once the results are extrapolated to the continuum. This is especially important
because in a walking or conformal model the system responds to a change in scale very mildly even over many orders of magnitude.
If one wishes to resolve this small continuum change
of the system in a lattice simulation at finite lattice spacing one needs to make sure that discretization errors are much smaller
than the small expected continuum change. 

In this review three approaches to studying the infrared properties of the $N_f = 2$ sextet model are outlined. At present none of
the conclusions from the three approaches are extrapolated reliably to the continuum. As we will see currently there are
disagreements between the simulation results but it might turn out that they are entirely due to discretization errors.

\section{Methods}
\label{methods}

There are many observables that are sensitive to the low energy dynamics of the model. The physics of chiral symmetry breaking is
very different from unbroken conformal symmetry and there are many ways to characterize this distinction. 

\subsection{Running coupling and step scaling}

The continuum renormalized coupling in the infrared distinguishes directly between the two cases of interest: if chiral symmetry breaking takes
place the coupling keeps growing towards the infrared whereas if conformal and chiral symmetry is intact the coupling
approaches a fixed point. Different coupling schemes might differ in the value $g_*^2$ of the fixed point but the fact that a fixed point
exists does not depend on the scheme. Hence any non-perturbatively well-defined scheme can be used and any physical scale can play
the role of the running scale $\mu$.

In lattice calculations a popular choice is the Schroedinger functional scheme \cite{Luscher:1992an}. In this case $\mu = 1/L$ where
$L$ is the finite size of the box. Instead of computing the $\beta$-function as the response of the coupling to an infinitesimal
scale change as in the continuum one typically calculates a discrete version on the lattice. This discrete $\beta$-function or
step scaling function measures the response of the coupling to a finite change, say a factor of $2$, in the scale. An infrared fixed point is both a
zero of the ordinary $\beta$-function and the discrete step scaling function. As in any lattice calculation a careful continuum
extrapolation is needed, the step scaling function can be zero at finite lattice spacing but this artifact fixed point can
disappear for smaller lattice spacing. In this context one lattice spacing refers to one lattice size change, say $N^4 \to
(2N)^4$, where $N = L/a$.
Changing the bare coupling $\beta$ is then used to change the renormalized coupling and the continuum limit corresponds to $N \to \infty$.
The continuum limit is especially important because often times the calculations are pushed to large bare coupling in order to see
traces of a possible fixed point exactly where lattice artifacts are large.

\begin{figure}
\begin{center}
\includegraphics[width=6cm,height=5.8cm]{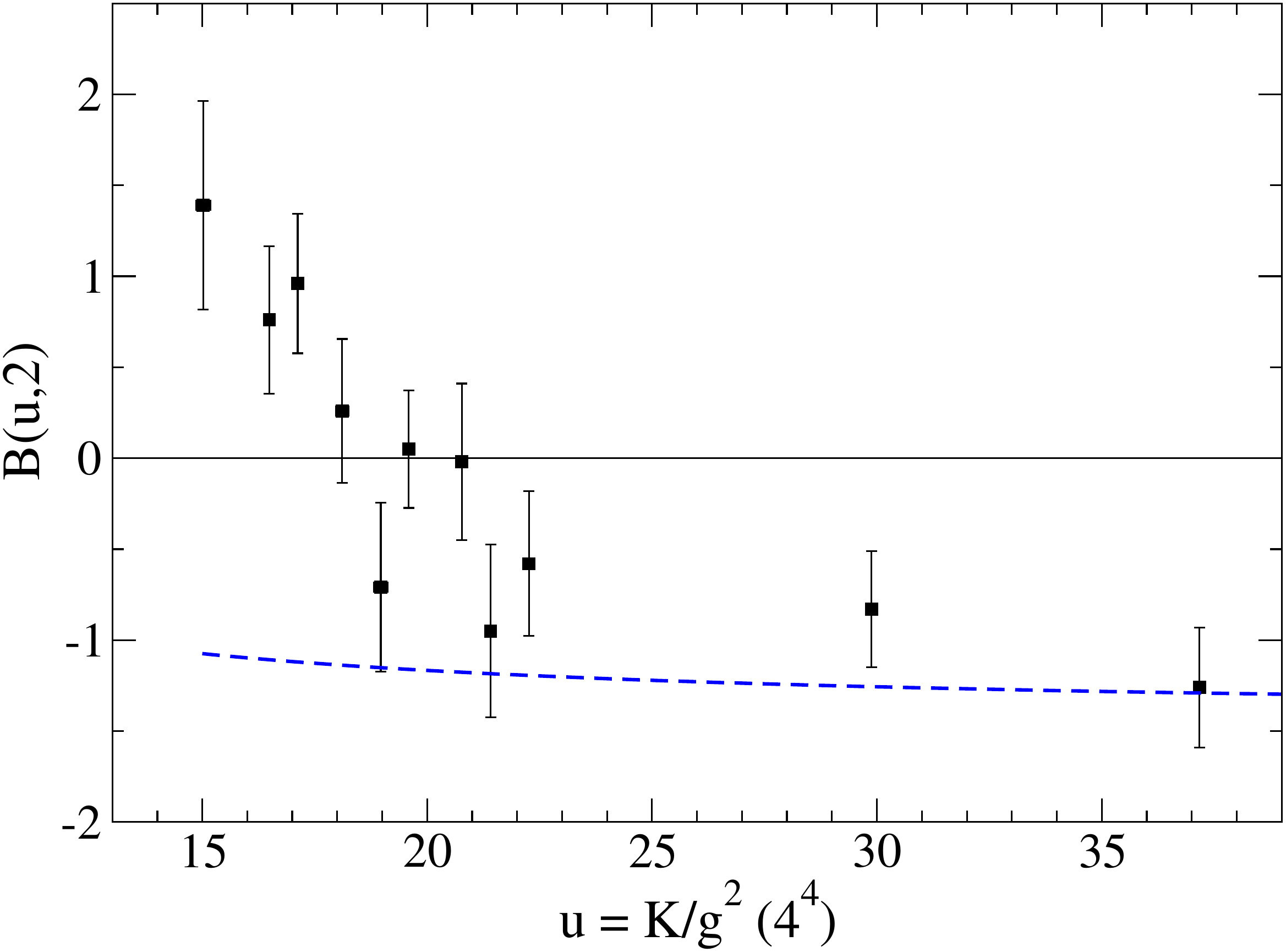} \hspace{1cm} \includegraphics[width=6cm]{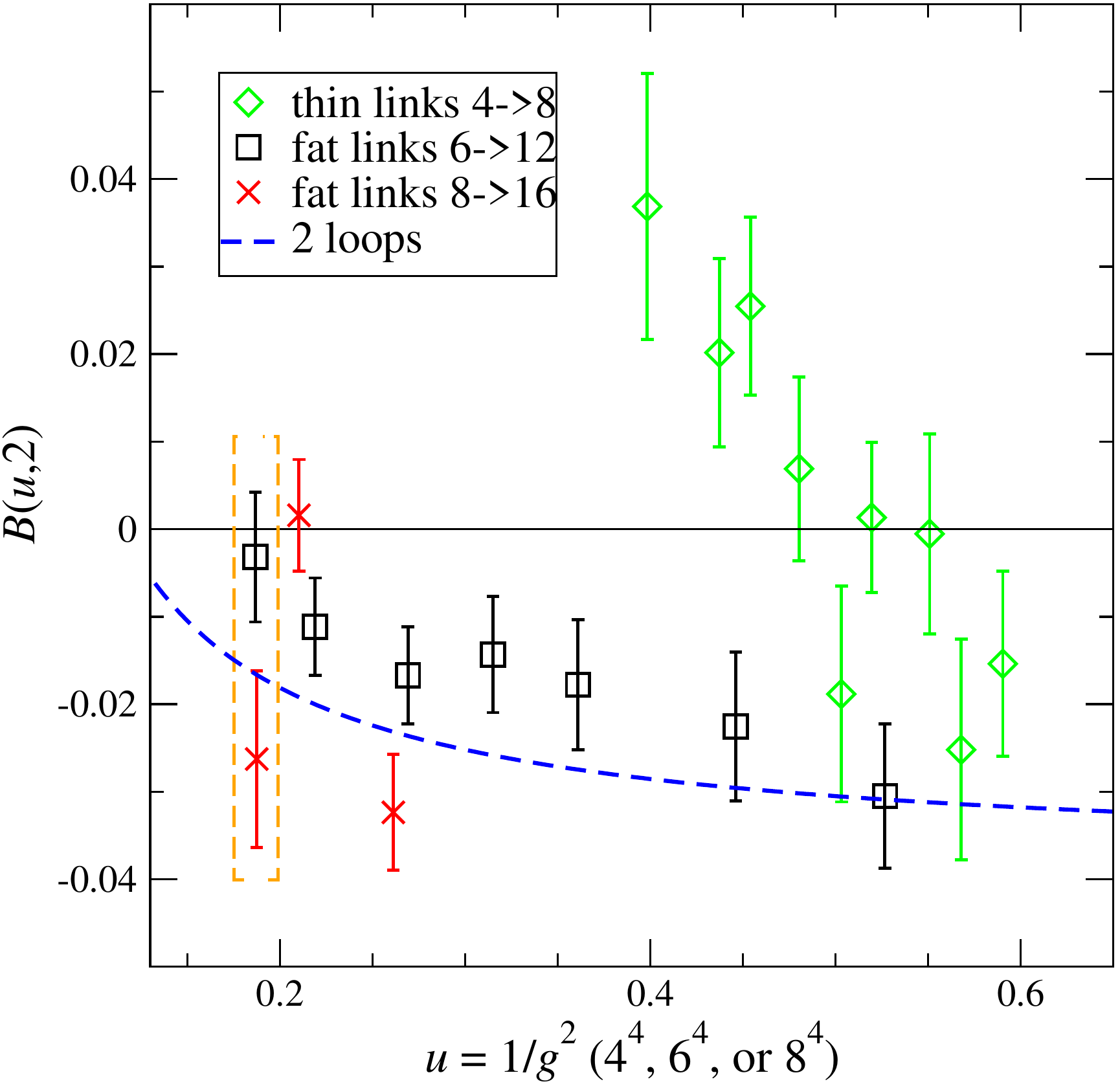}
\end{center}
\caption{The step scaling function calculated in \cite{Shamir:2008pb} (left) with thin links indicating an infrared fixed point. Using fat links
for the fermion action (right) the fixed point disappears \cite{DeGrand:2010na}. See the text for more details.}
\label{tom1}
\end{figure}

\begin{figure}
\begin{center}
\includegraphics[width=8cm,height=6cm]{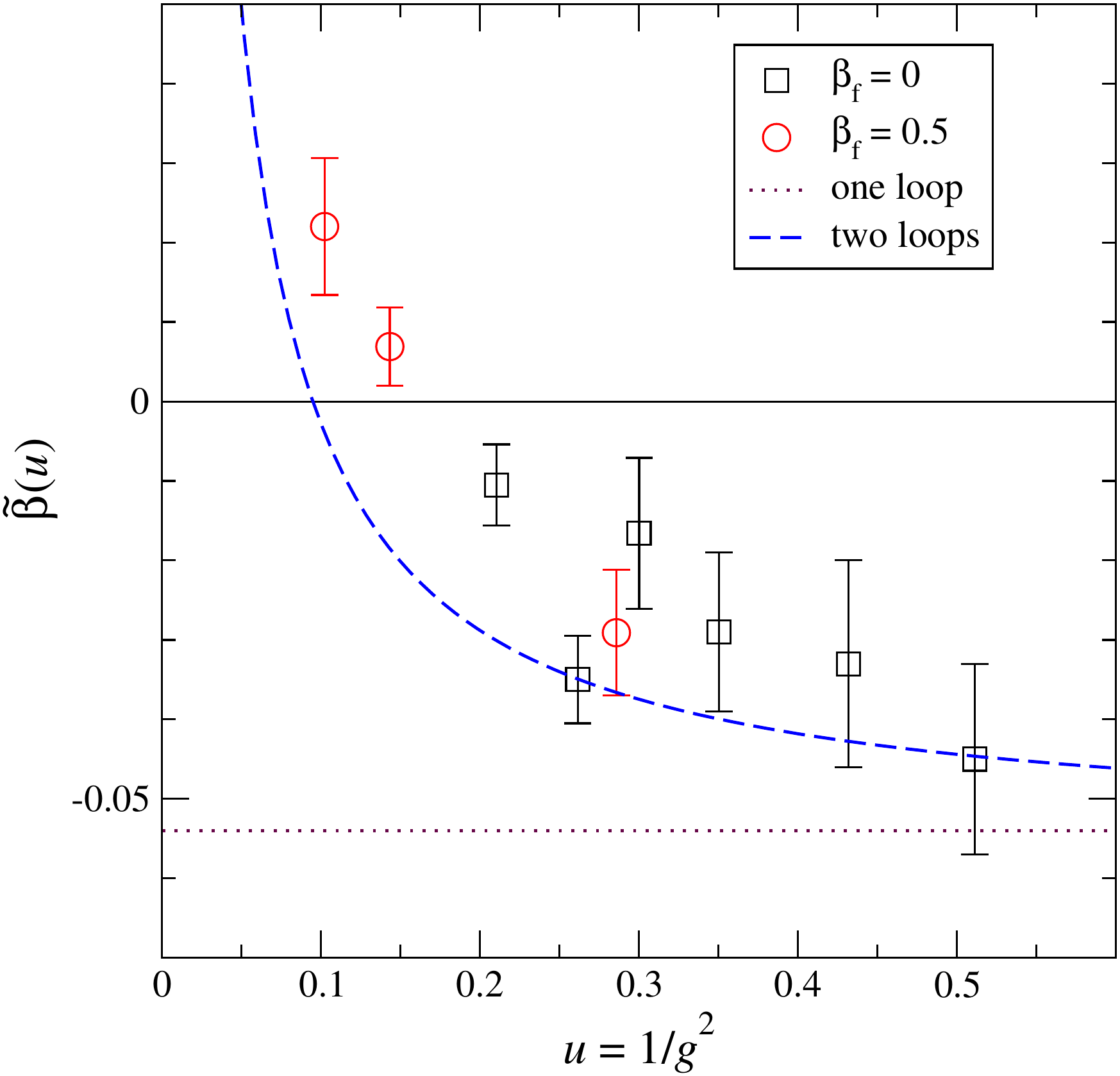}
\end{center}
\caption{The step scaling function from \cite{DeGrand:2012yq} using fat links for the fermion action only (blue) and fat links for both the
fermion and gauge actions (black). The fixed point is visible again; see the text for more details.}
\label{tom2}
\end{figure}

The calculation of the running coupling in the Schroedinger functional scheme using Wilson fermions was started in \cite{Shamir:2008pb} for the
$N_f = 2$ sextet model. Using
an unimproved (thin link) Wilson fermion action a zero of the step scaling function was measured at one lattice spacing corresponding to
$4^4 \to 8^4$, see left panel of figure \ref{tom1}. Two more lattice spacings corresponding to $6^4\to 12^4$ and $8^4\to 16^4$ 
were then added \cite{DeGrand:2010na} using an improved (fat link) Wilson fermion
action, see right panel of figure \ref{tom1}. The fixed point disappeared with a possible interpretation that the rougher lattice
spacing result was an artifact. The gauge action was the same in the two calculations. However changing not only the fermion action
but the gauge action as well to use fat links resulted in a step scaling function with a zero for the lattice spacing corresponding to $6^4\to
12^4$, see figure \ref{tom2}. A possible interpretation is that the absence of the zero previously was the artifact after all
\cite{DeGrand:2012yq}. 

Changing the action and/or the lattice spacing led to results so far which show that discretization effects are still there. 
Clearly a careful continuum extrapolation is necessary with a given action in order to decide which finite lattice spacing result is
the one prevailing all the way to the continuum. A good check of the procedure would be the reproduction of the 2-loop
$\beta$-function for small renormalized coupling, carefully extrapolated to the continuum. 

As a cross-check it would be helpful if the running coupling would be calculated in a different non-perturbatively well-defined
scheme. Reproducing the 2-loop $\beta$-function for small coupling is always a good test for any scheme. For larger coupling two schemes can
disagree on the value of the coupling but if a fixed point exists for one scheme a fixed point should exist for the other scheme
too.

\subsection{Thermodynamics}

Another way of addressing the infrared behavior of the model is studying it at finite temperature. If chiral symmetry is broken
at $T=0$ one expects a chiral symmetry restoration temperature $T_c$. If the model is conformal in the infrared then as far as
chiral symmetry is concerned there is no phase transition at all for $T>0$. Lattice investigations of thermodynamical properties are
complicated by the fact that the lattice system at finite lattice spacing typically has a rich phase structure with various types of phase transitions and
phase boundaries most of which however happens to be regularization specific and as such an artifact with no consequence
to the continuum. Bulk phase transitions are an example. A careful continuum extrapolation of the findings is hence again
essential.

The thermodynamic study of the $N_f = 2$ sextet model was initiated in \cite{Kogut:2010cz}. Using unimproved rooted staggered fermions in the
fixed$-N_t$ approach the Polyakov loop, the chiral condensate and chiral susceptibility were measured at various quark masses. In the fixed$-N_t$ approach
one lattice spacing corresponds to a given $N_t = 1/(aT)$, $\beta$ is used to change the temperature and the continuum 
limit is achieved via $N_t \to \infty$. A thermal phase 
transition corresponds to a critical $\beta_c(N_t)$ coupling for each $N_t$ which for large $N_t$ scales according to the
continuum $\beta$-function; in particular $\beta_c \to \infty$. A bulk phase transition on the other hand is 
characterized by critical $\beta_c(N_t)$ couplings which do not scale and for large $N_t$ approach a fixed value.

\begin{figure}
\begin{center}
\includegraphics[width=8cm]{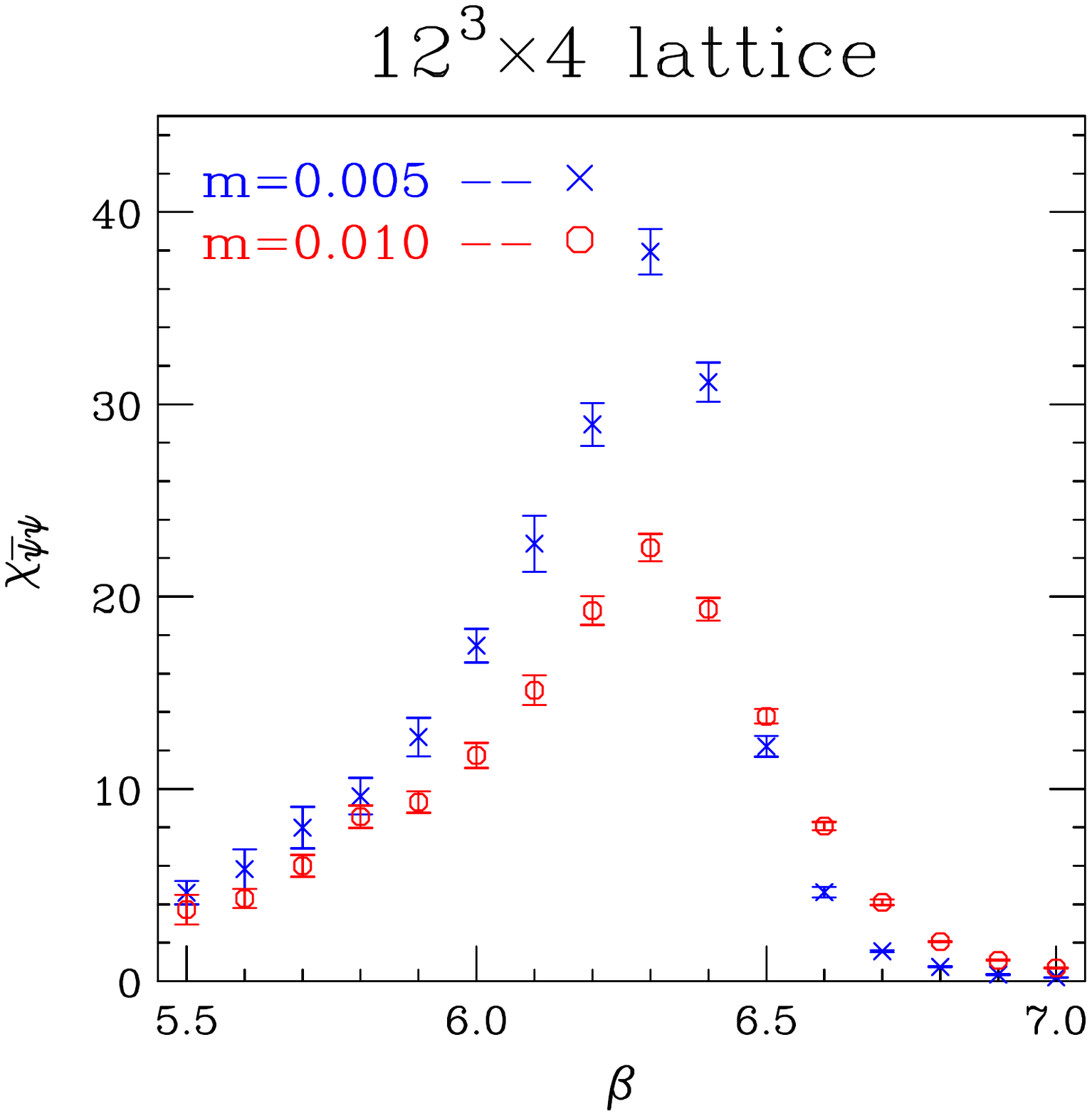} \hspace{-1.5cm} \includegraphics[width=8cm]{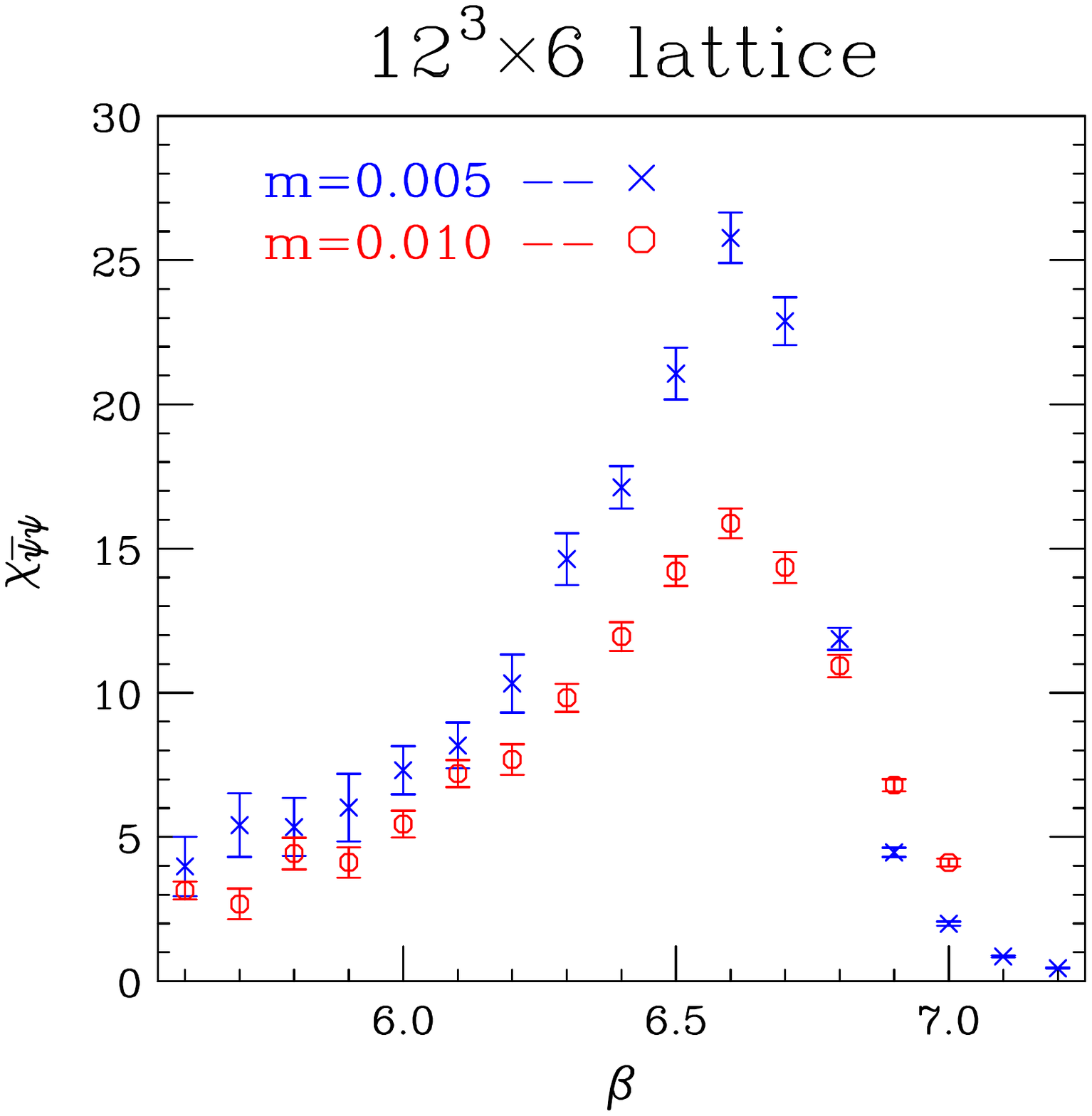} \\
\vspace{-2cm}
\includegraphics[width=8cm]{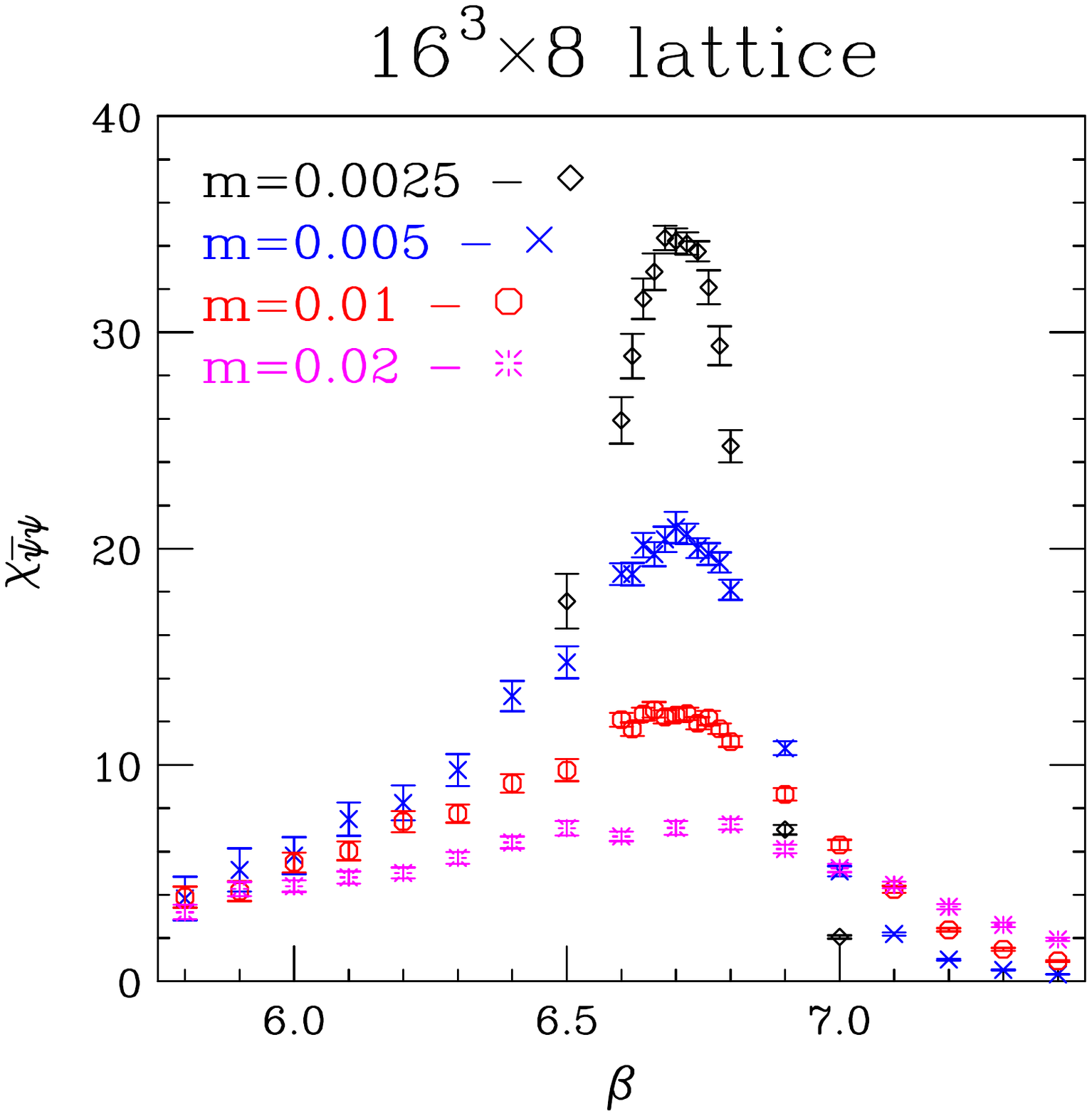} \hspace{-1.5cm} \includegraphics[width=8cm]{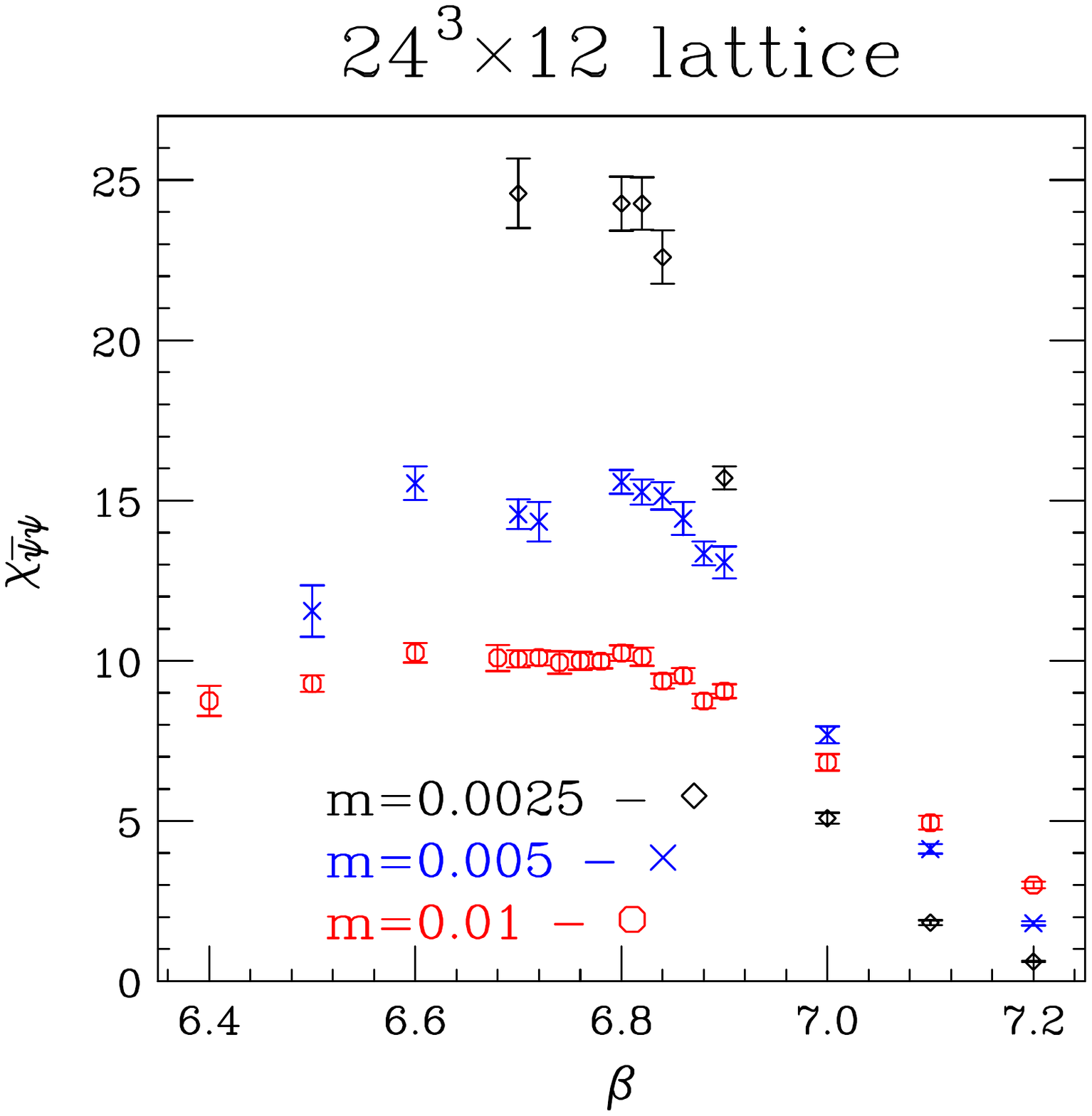}
\vspace{-2cm}
\end{center}
\caption{The chiral susceptibility on $N_t = 4$ and $N_t = 6$ lattices (top) from \cite{Kogut:2010cz},
and on $N_t = 8$ and $N_t = 12$ lattices from \cite{Kogut:2011ty} and \cite{Sinclair:2011ie} respectively (bottom).}
\label{don}
\end{figure}

As always with any thermodynamics study finite volume effects need to be under control and the quark mass needs to be small enough.
Since staggered fermions are used the lattice spacing also needs to be small enough in order to avoid dangerous taste violation effects
especially because the low energy dynamics is very sensitive to the number of massless flavors.  

The critical coupling $\beta_c$ was determined in \cite{Kogut:2010cz} from the peak of the chiral susceptibility on $N_t = 4$
lattices for two values of the quark mass. The location of the peaks appear to be mass
independent and is around $\beta_c \approx 6.3$, see top left panel of figure \ref{don}. The $N_t = 6$ result at the same two quark
masses also from \cite{Kogut:2010cz} is shown on the top right panel of figure \ref{don}. The critical coupling moved to $\beta_c \approx 6.6$. On even
finer lattices \cite{Kogut:2011ty}, at $N_t = 8$, the critical coupling moved further, to around $\beta_c = 6.7$ with additional small quark masses
added, see bottom left panel of figure \ref{don}. Again the quark mass dependence is quite small. Finally the $N_t = 12$ lattices are
preliminary \cite{Sinclair:2011ie} at the moment but seem to indicate further increase in $\beta_c$, see the bottom right panel of figure \ref{don}. If indeed
$\beta_c$ scales with $N_t$ correctly the located phase transitions would correspond to a continuum phase transition indicating
chirally broken symmetry at zero temperature.

A priori it is not clear how large $N_t$ needs to be in order to be in the scaling regime.
Most importantly the thin link action suffers from possible large taste violation. Unfortunately, these effects are not
quantified yet. One could in principle reduce them by using smeared actions. In any case a continuum extrapolation is necessary.

\subsection{Meson spectrum}
\label{mesonspectrum}

Yet another way of deciding whether the model in the infrared is conformal or chirally broken is via studying the mass spectrum
as a function of the quark mass. If chiral symmetry is broken chiral perturbation theory \cite{Gasser:1983yg} provides a quantitative description of
the quark mass dependence of low energy observables. These typically contain polynomial and logarithmic terms. The leading
formulae are followed by sub-leading terms and it is not clear how small the quark mass needs to be in order for the
sub-leading terms to be suppressed sufficiently. The order in the chiral expansion which needs to be taken into account may also
depend on the observable as well. The functional form of the leading and sub-leading terms is however known. 

The application of infinite volume chiral perturbation theory requires of course that the finite volume of the system is large
enough so that increasing it further does not change the measured observables. Finite volume effects are exponential and controlled by $m_\pi L$
as usual but how large this needs to be is model dependent. Models expected to be close to the conformal window
typically display much larger finite volume effects than QCD and $m_\pi L > 8 - 10$ is sometimes necessary. The applicability of
infinite volume chiral perturbation theory also requires a large $f_\pi L$ value but again it is not clear how large it needs to
be and the bound probably depends on the observable. Small enough lattice spacing is again essential if staggered fermions are used because of taste violation.

If the model is conformal in the infrared the leading quark mass dependence is power-like and fixed by the mass anomalous
dimension, $M(m) \sim m^{1/(1+\gamma)}$. This scaling holds for all masses and decay constants as well \cite{DelDebbio:2010ze}. For the chiral condensate
one has ${\bar\psi}\psi \sim m + c m^{(3-\gamma)/(1+\gamma)}$.
Again it is not known a priori how small the quark masses need to be in order to neglect sub-leading terms.
The functional form of the sub-leading terms is however not known. The scaling formula holds in infinite volume and just as with the
chirally broken case finite volume effects are needed to be small in order to apply them. Finite volume effects are again
exponential in $m_\pi L$ where now $m_\pi$ simply means the mass of a particle with the same quantum numbers as the pion of QCD. 
The finite size of the system can also be
used to study finite size scaling and investigate whether a consistent scaling variable can be built out of the quark mass $m$ and
$L$. 

The meson spectrum was first studied in \cite{Fodor:2011tw} using stout improved rooted staggered fermions and Symanzik gauge
action. The bare quark mass was varied in the range $0.003 - 0.014$, the gauge coupling was set to $\beta = 3.2$ while the
lattice volume was $24^3 \times 48$ and $32^3 \times 64$. The finite volume effects were not fully explored in \cite{Fodor:2011tw}
and the measured observables on $32^3 \times 64$ were interpreted as infinite volume results. This assumption was subsequently
justified in \cite{Fodor:2012uu}; see below.

\begin{figure}
\begin{center}
\includegraphics[width=7cm]{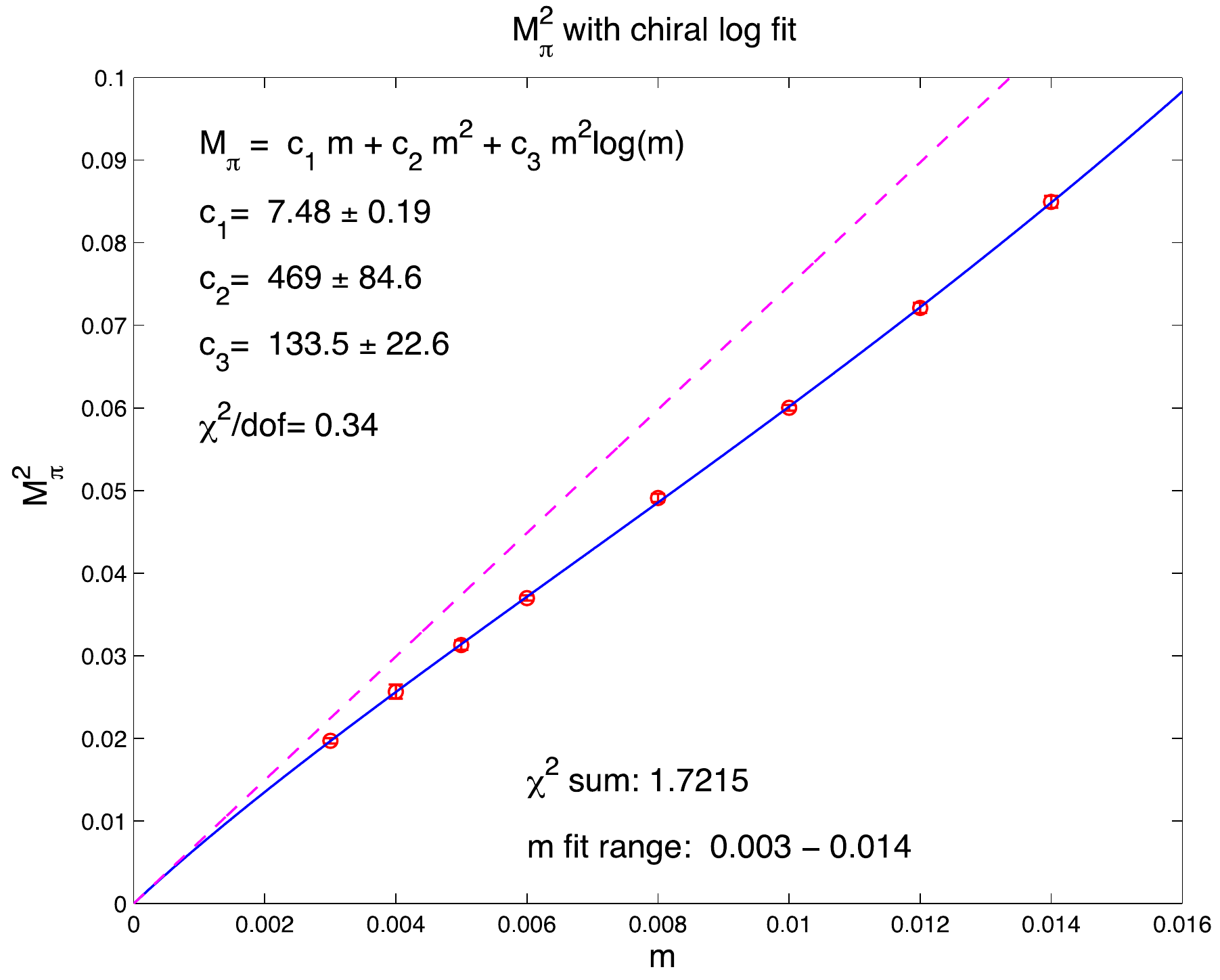} \includegraphics[width=7cm]{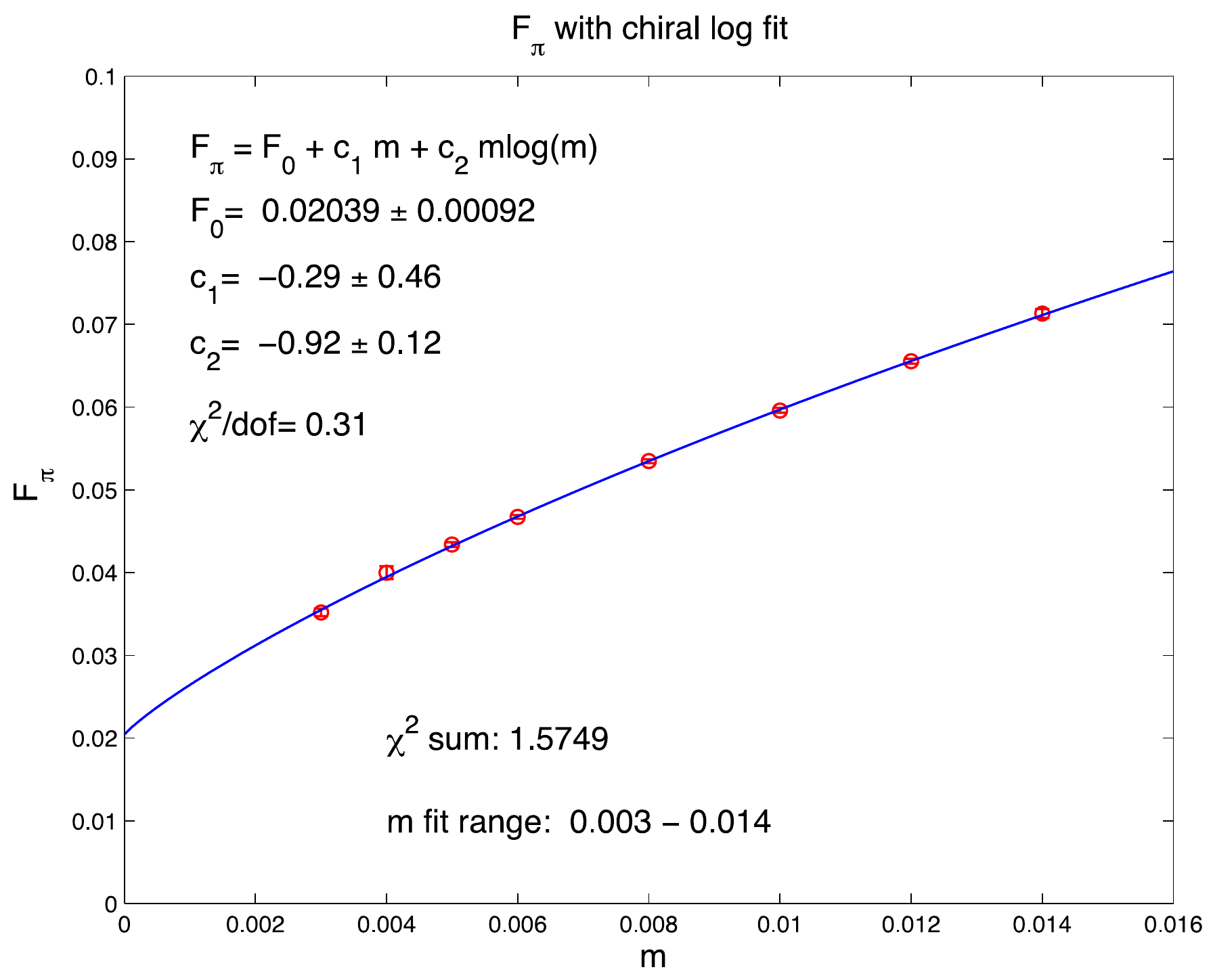} \includegraphics[width=7cm]{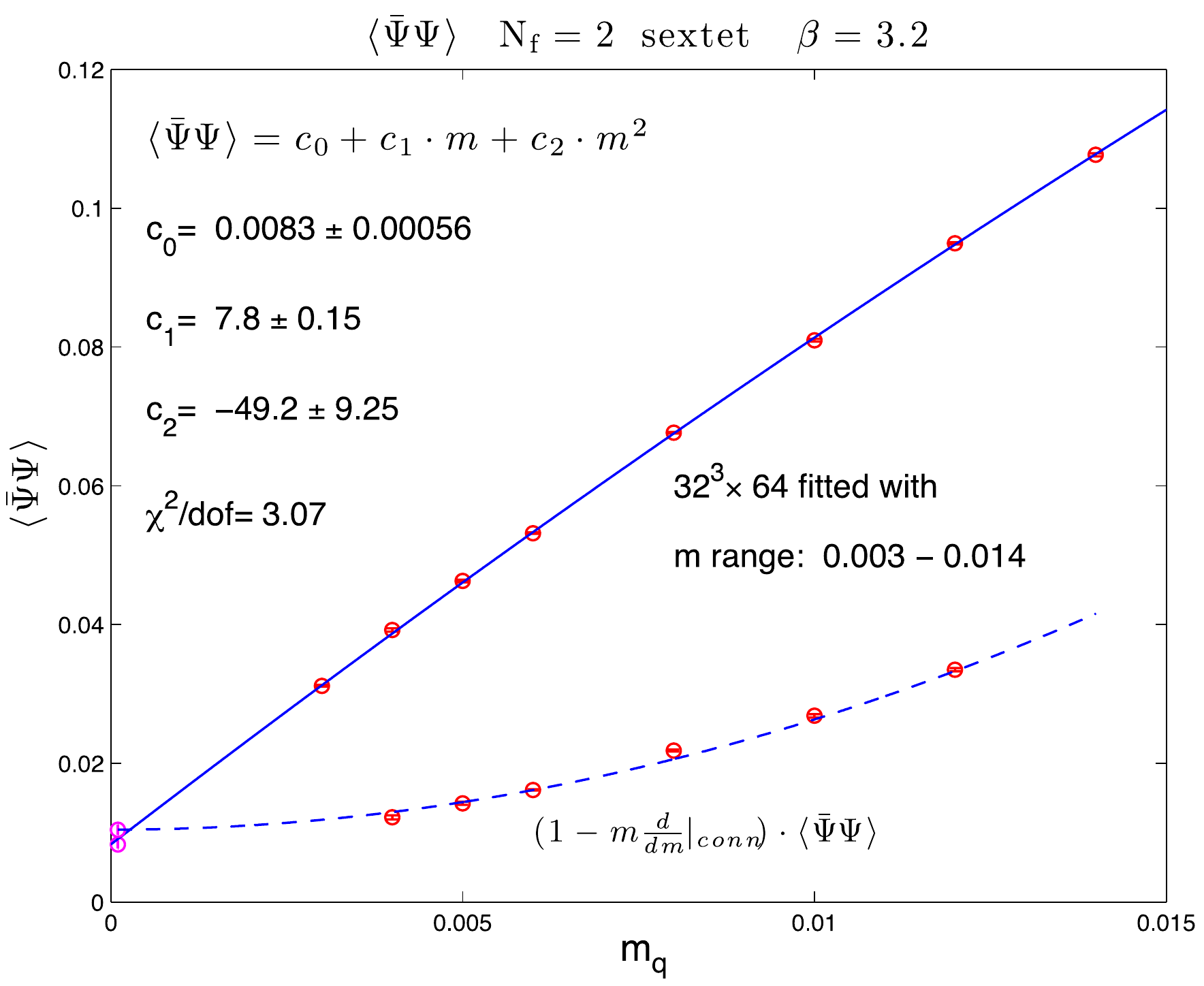}
\end{center}
\caption{1-loop chiral perturbation theory fits to the pion mass (top left), decay constant (top right) and chiral condensate
(bottom) from \cite{Fodor:2011tw}. For the pion mass the leading order result is also shown in pink.}
\label{chiralpert}
\end{figure}

The 1-loop infinite volume chiral perturbation theory fits for the pion mass, decay constant and chiral condensate are shown on
figure \ref{chiralpert} from \cite{Fodor:2011tw}. The chiral condensate itself is shown and also an independently measured quantity which subtracts the term linear
in $m$ using the chiral susceptibility. The chiral limit value is of course consistent between the two determinations as it should
be. The advantage of using the subtracted condensate is that this quantity changes much less over the fitted mass range than the
condensate itself making the extrapolation to the chiral limit more reliable. 

It is clear from figure \ref{chiralpert} that both the condensate and the decay constant extrapolate to non-zero values in the chiral
limit, consistently with spontaneous chiral symmetry breaking in the massless theory.

In order to test the compatibility or incompatibility with a conformal massless theory, leading order conformal fits using
power-like forms were also performed. The resulting global $\chi^2/{\rm dof}$ values for
the two fits are $1.24$ for the chirally broken hypothesis using 1-loop chiral perturbation theory and $6.96$ for the conformal
hypothesis. The global fit in both cases used
the same 3 channels, pion mass, decay constant and chiral condensate. Clearly the analysis in \cite{Fodor:2011tw} 
favors the chirally broken hypothesis over the conformal one. However it is doubtful whether 1-loop chiral perturbation theory is
trustworthy in the full fermion mass range for the decay constant; see below.

\begin{figure}
\begin{center}
\includegraphics[height=6cm]{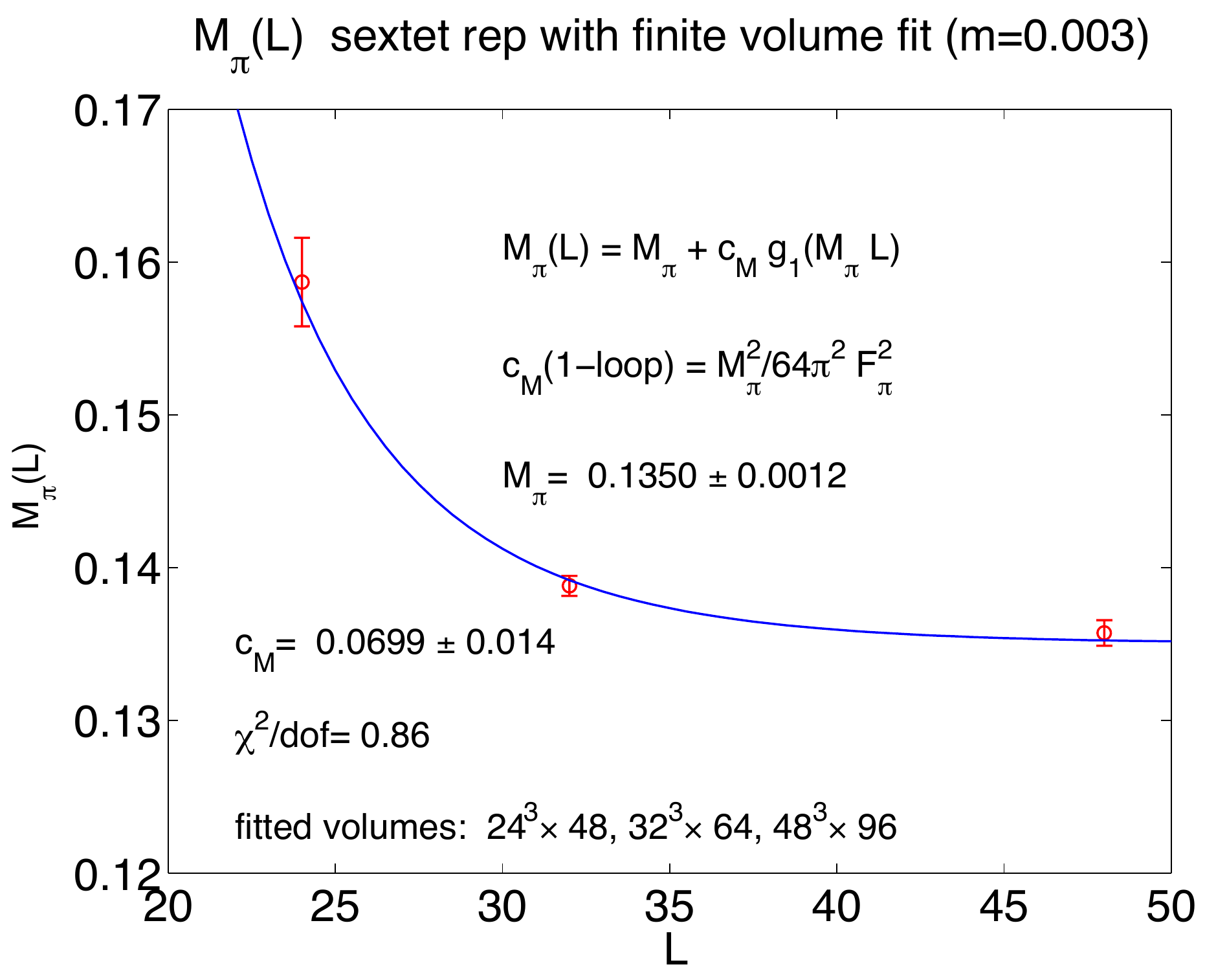} 
\includegraphics[height=6cm]{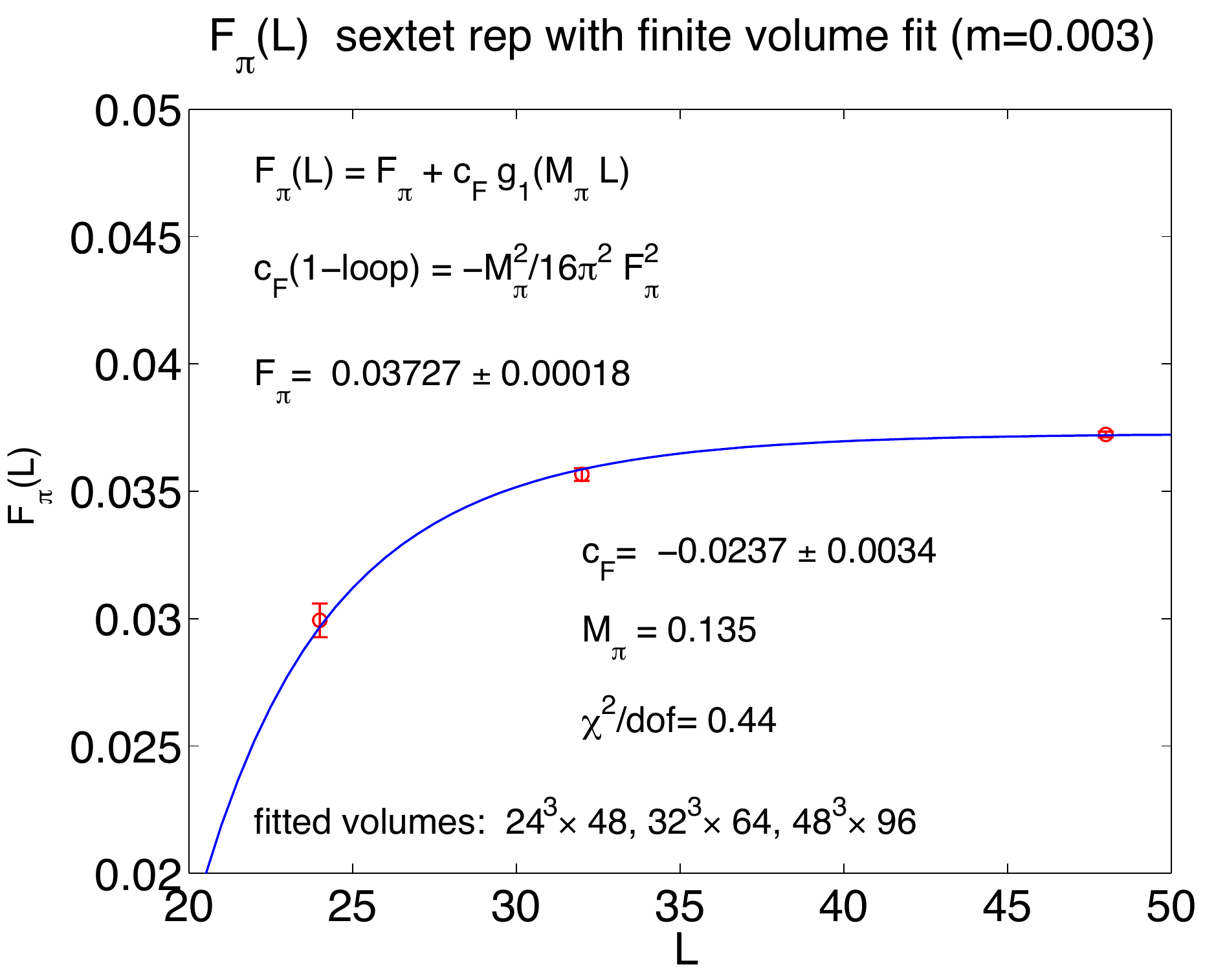}
\includegraphics[height=6cm]{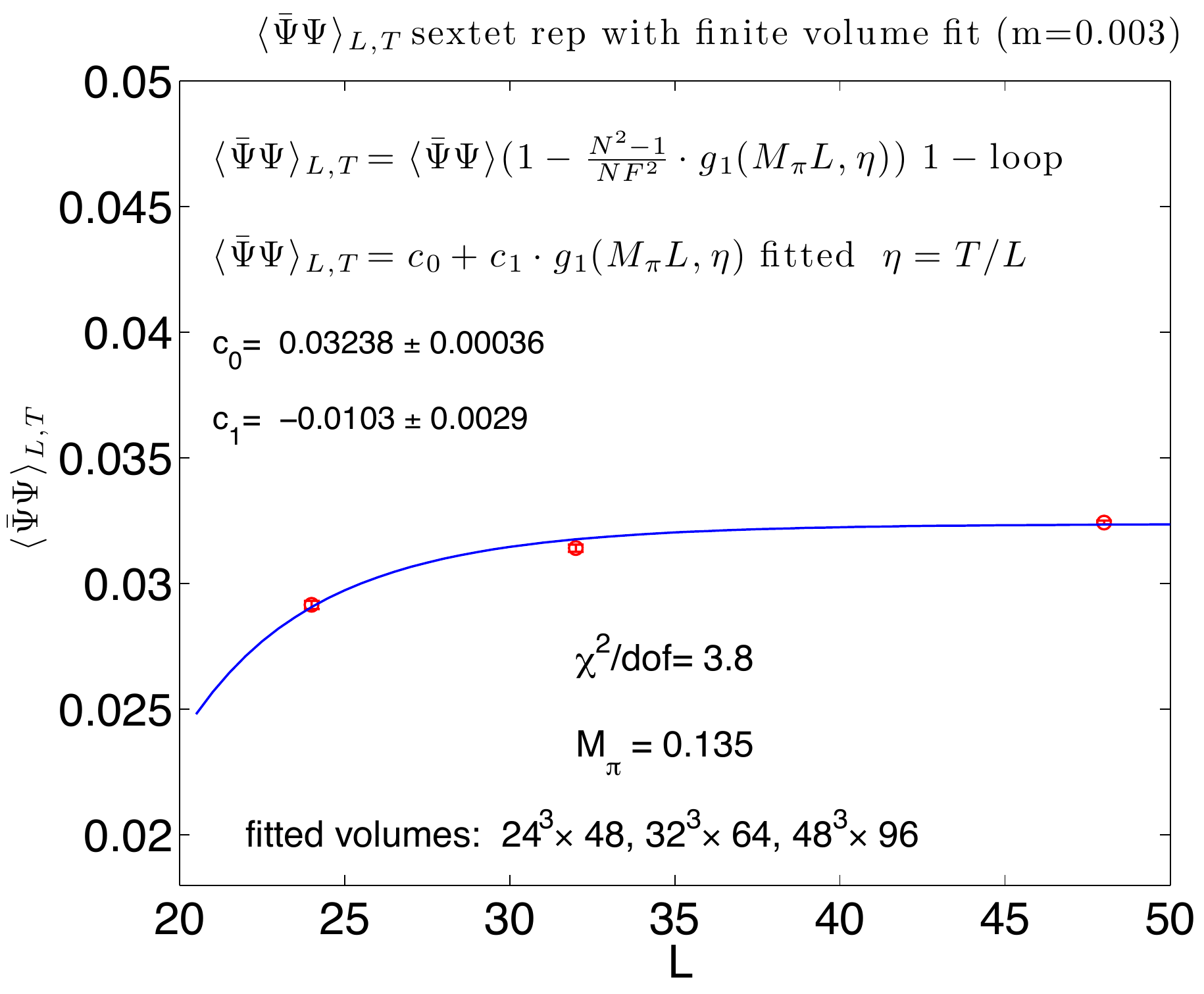}
\end{center}
\caption{Volume dependence of the pion mass (top left), decay constant (top right) and chiral condensate (bottom) for $m = 0.003$,
$\beta = 3.20$ and volumes $24^3 \times 48$, $32^3 \times 64$ and $48^3 \times 96$ from \cite{Fodor:2012uu}.}
\label{finitel}
\end{figure}

As mentioned already the applicability of infinite volume formulae for both the chirally broken and conformal scenarios requires
large enough lattices. This aspect of the analysis in \cite{Fodor:2011tw} was investigated in \cite{Fodor:2012uu} where additional
simulations at $\beta = 3.2$ and $m =0.003$ were performed on $48^3 \times 96$ lattice volumes. 
At this mass the pion mass, decay constant and the chiral condensate are shown on figure
\ref{finitel} together with (an essentially exponential) fit to the volume dependence.
As can be seen the difference between the $32^3$ and $48^3$ spatial volumes is very small hence will be even
smaller for $m > 0.003$. Consequently in the quark mass dependence fits the $32^3$ volumes can indeed be used as infinite volume results
for all masses. Figure \ref{finitel} leads to the requirement $m_\pi L > 5$ in order to have essentially no further volume
dependence (below 1\% level).

Another shortcoming of the analysis in \cite{Fodor:2011tw} was also addressed in \cite{Fodor:2012uu}.
It is clear from figure \ref{chiralpert} that the effect of the subleading term in the chiral expansion is quite small for the pion mass and
subtracted chiral condensate. These two quantities are most probably extrapolated to the chiral limit reliably. However
the decay constant changes quite a bit over the fitted mass range all the way to the chiral
limit. The difference between the leading order (constant) and subleading (logarithm) terms in the chiral expansion is substantial
raising the question whether higher order corrections need to be taken into account. The data in \cite{Fodor:2011tw} does not allow fitting 
2-loop chiral perturbation theory unfortunately. Another option would be to use the 1-loop formulae for a smaller mass range where
the deviation between the leading and subleading terms is smaller however this would leave too few data points for the analysis
and would not determine the unknown coefficients with any precision.

As mentioned already the mass range where the chiral logarithms become significant is a priori not known. The polynomial terms in chiral
perturbation theory can also be used to fit the data, which was performed in \cite{Fodor:2012uu} together with increasing the
statistics at each point. The mass range was set to the
smaller interval $m = 0.003 - 0.008$ in order to avoid higher masses where the applicability of any small mass expansion is
questionable. Since the results from the $48^3 \times 96$ volumes at $m = 0.003$ gave convincing evidence that the $32^3 \times 64$
results are quite close to the infinite volume limit, again the latter volumes were used in the fits for $m>0.003$ and the
infinite volume extrapolated result for $m = 0.003$; see figure \ref{finitel}.

\begin{figure}
\begin{center}
\includegraphics[height=6cm]{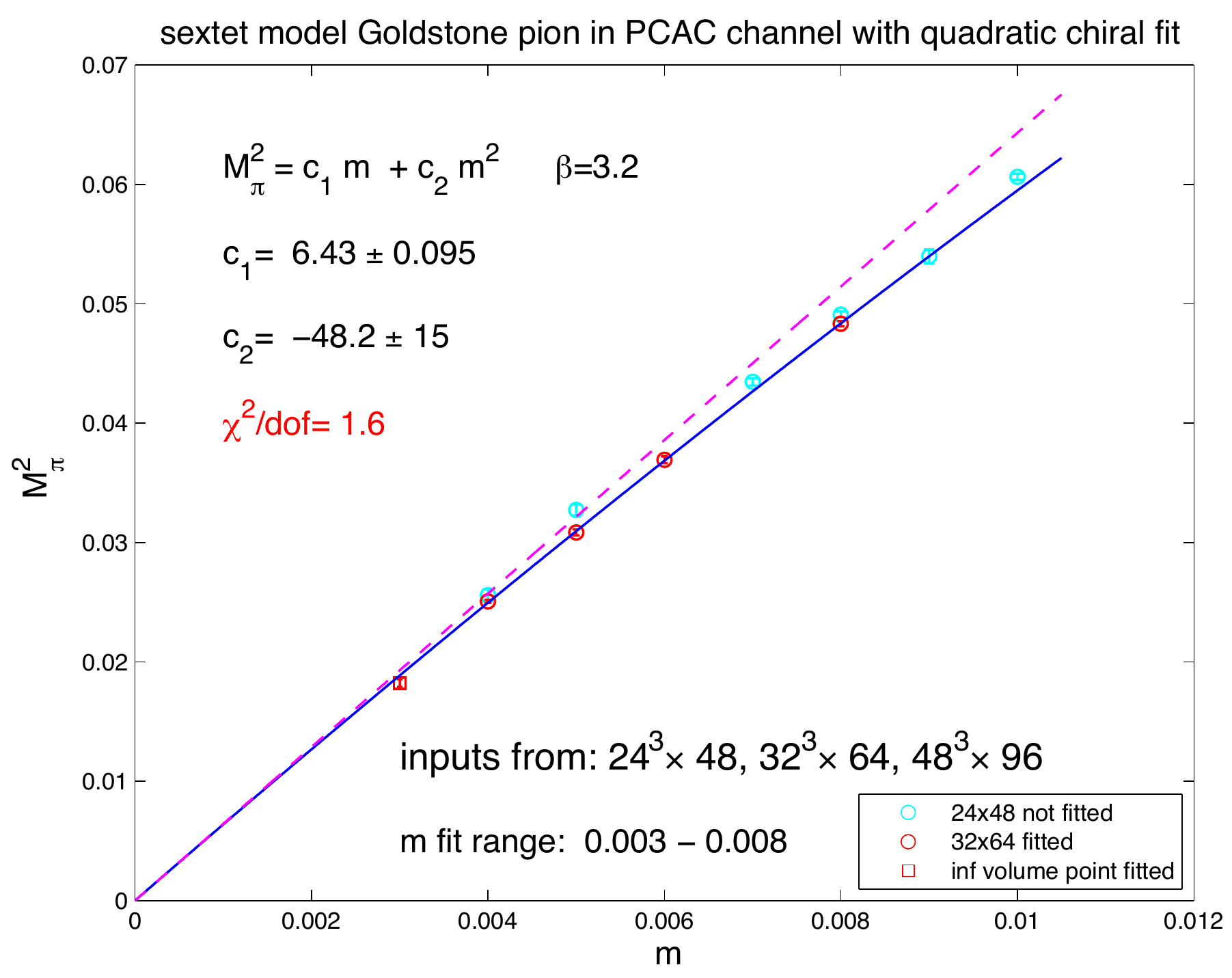}
\includegraphics[height=6cm]{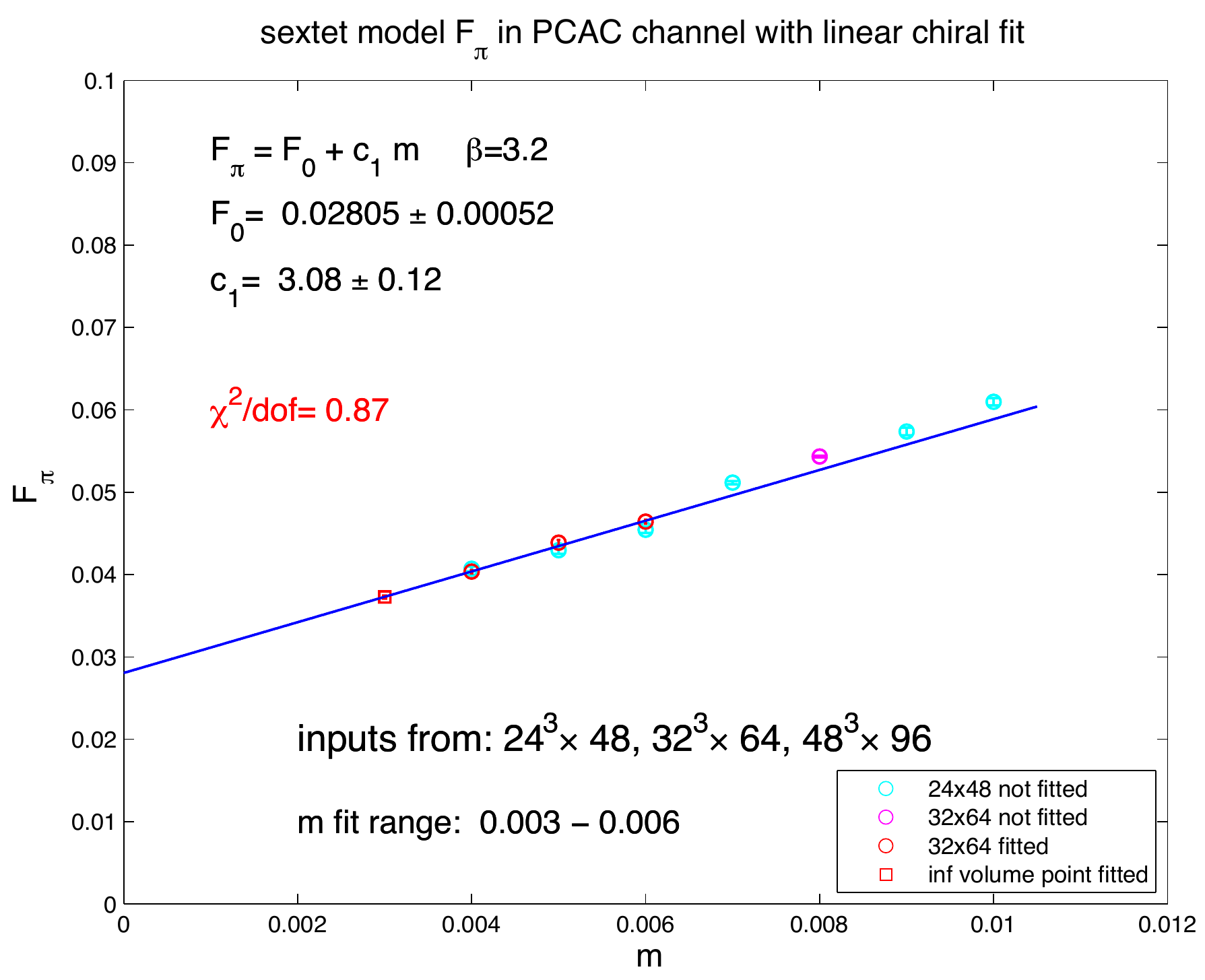} \\
\includegraphics[height=6cm]{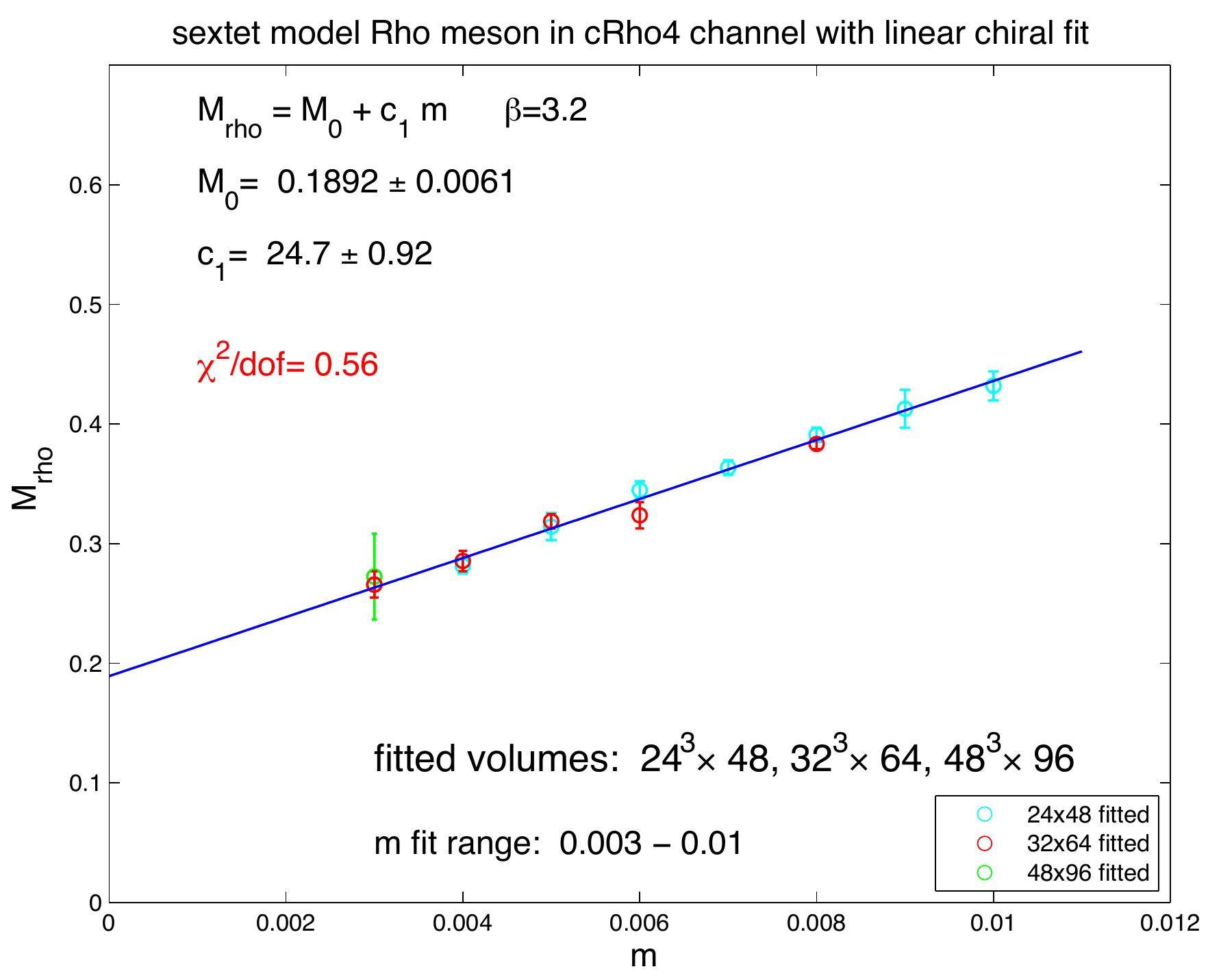}
\includegraphics[height=6cm]{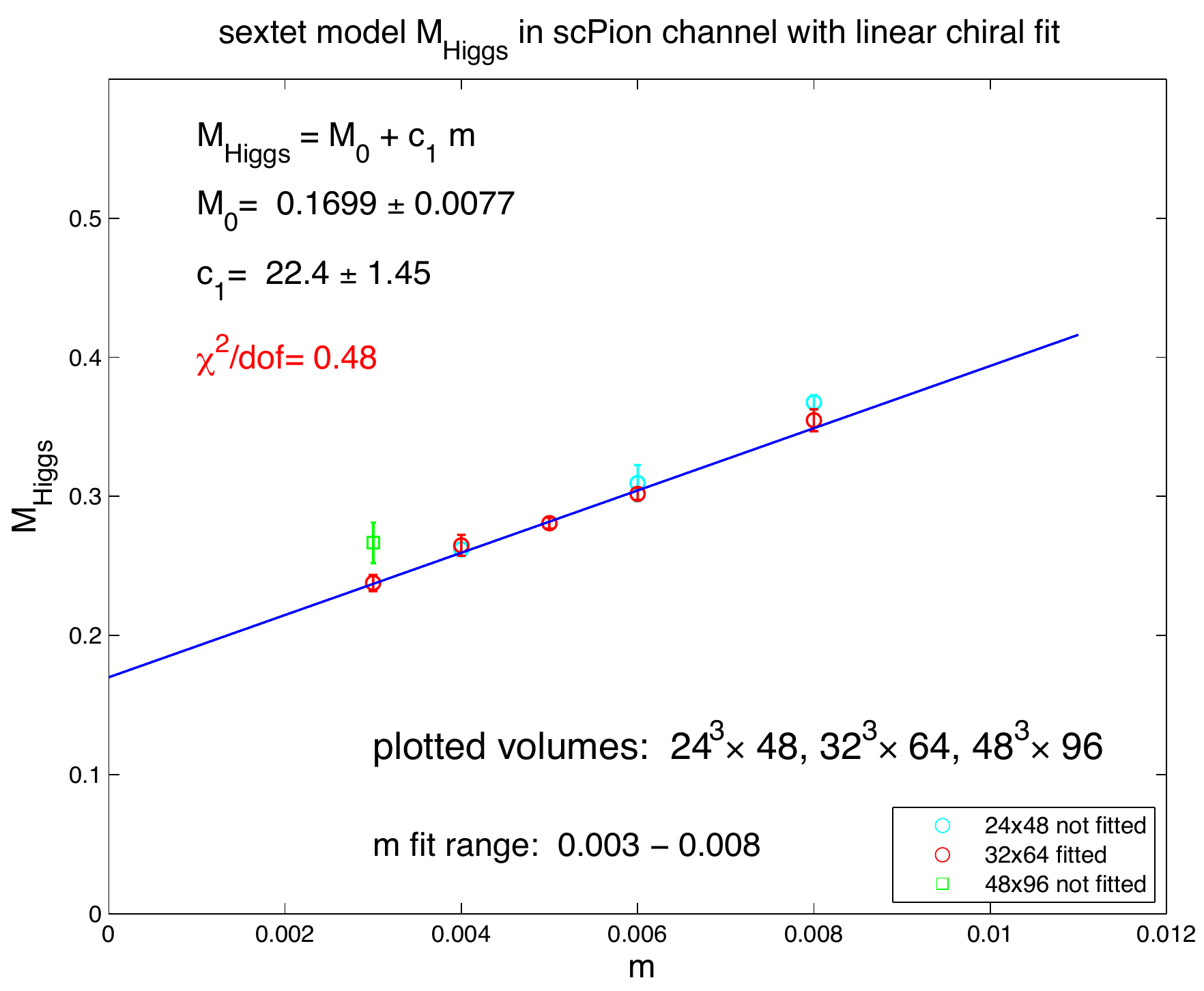}
\end{center}
\caption{Polynomial fits for the quark mass dependence of the pion mass (top left), decay constant (top right), $\varrho$-mass
(bottom left) and the composite Higgs $0^{++}$ mass (bottom right) from \cite{Fodor:2012uu}.}
\label{analytic}
\end{figure}

In figure \ref{analytic} polynomial mass dependence is fitted for the pion
mass, decay constant, $\varrho$ mass, and for what can be called the Higgs mass which is the $0^{++}$ particle or $\sigma$-particle (or
resonance) of QCD. In composite Higgs models, such as perhaps the $N_f = 2$ sextet model, this particle corresponds to the
composite Higgs. This last plot is incomplete in the sense that neither the disconnected propagator nor mixing with glueball
states has been measured and is shown only as illustration.
As shown on figure \ref{analytic} the fits have acceptable $\chi^2/{\rm dof}$ values. Taken together
with the non-vanishing values of the decay constant and chiral condensate they indicate that the data is compatible with
spontaneous chiral symmetry breaking and hence non-conformal infrared behavior.

\begin{figure}
\begin{center}
\includegraphics[height=6cm]{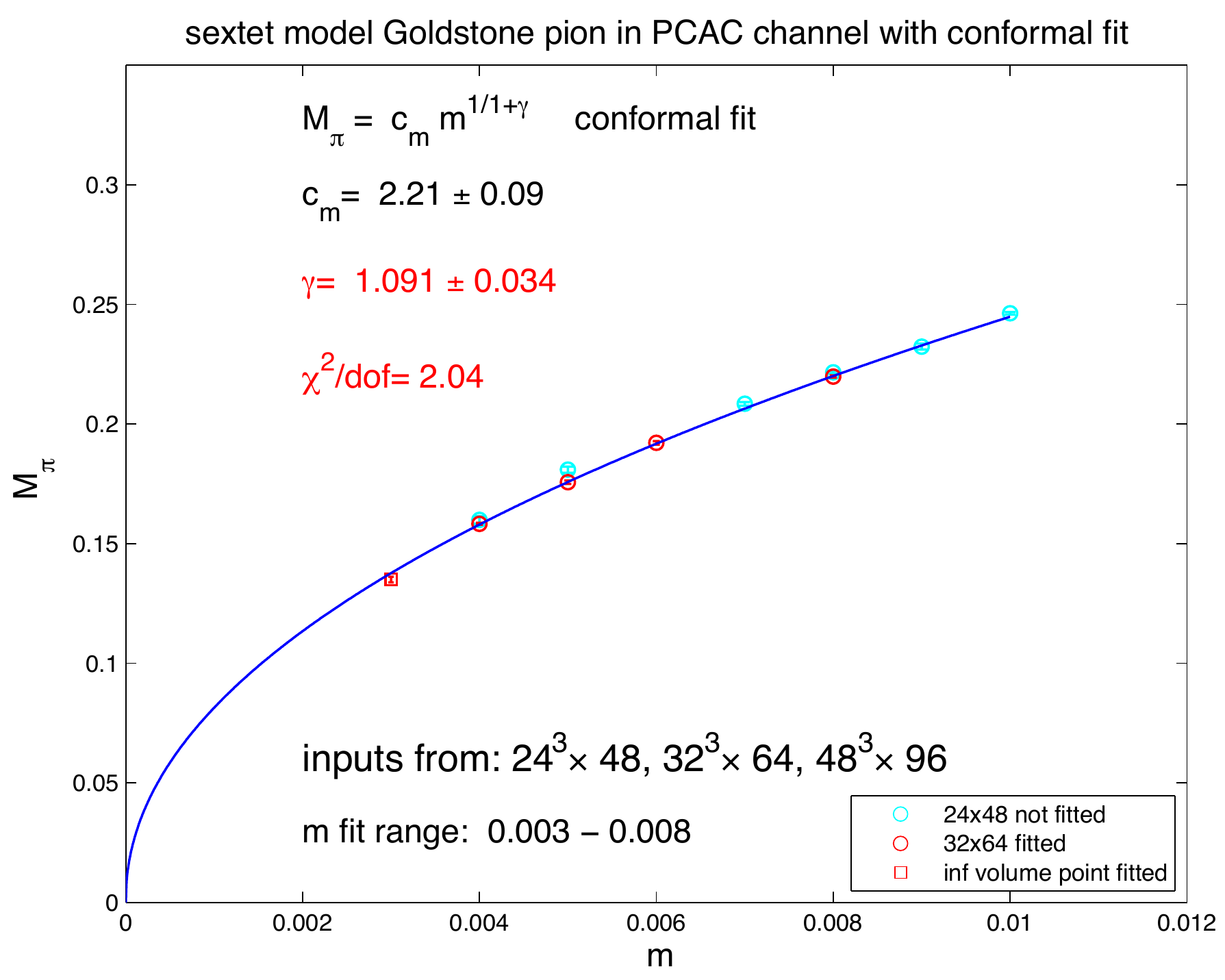} \includegraphics[height=6cm]{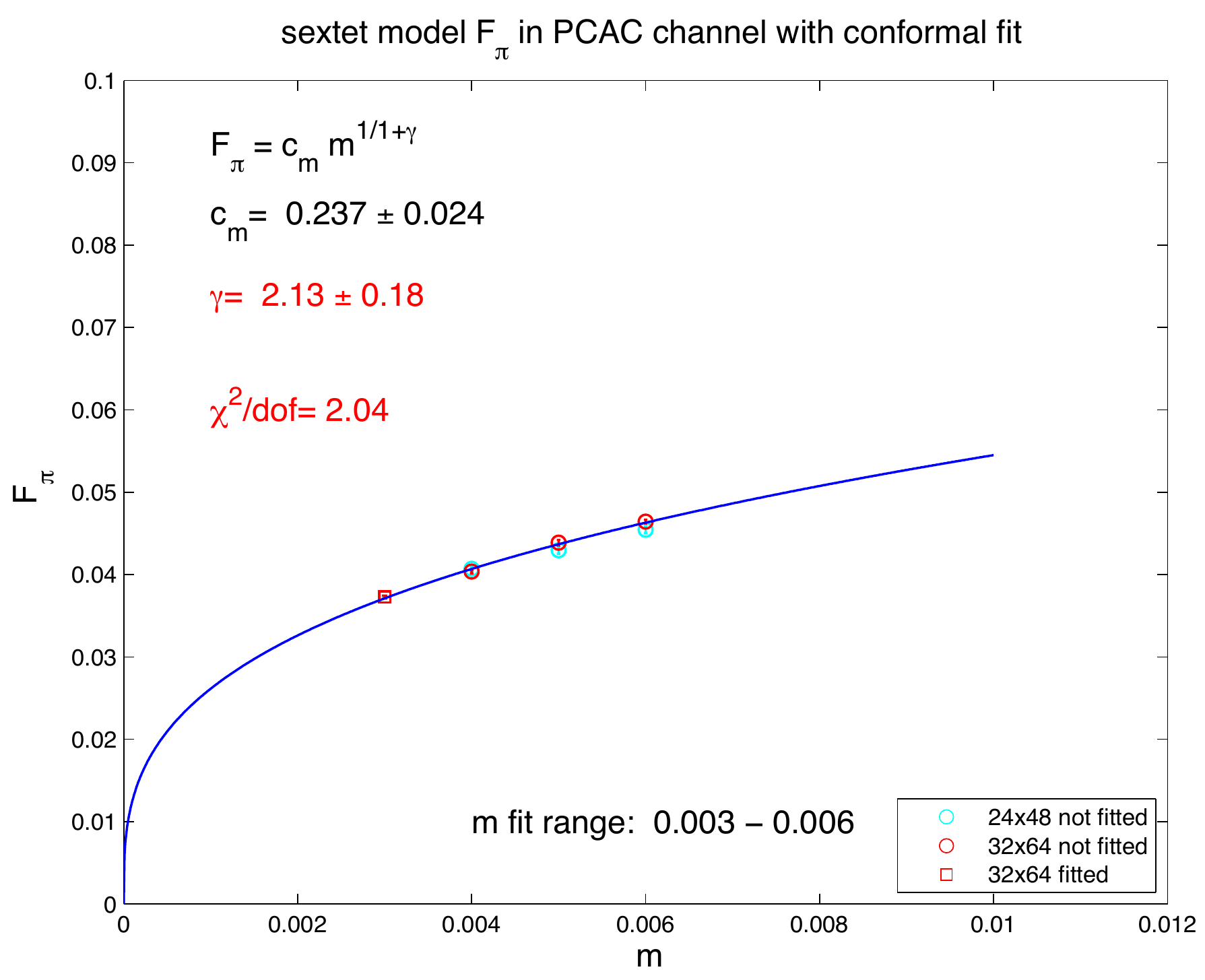}
\end{center}
\caption{Conformal fit to the decay constant (left) and pion mass (right) using separate $\gamma$ anomalous dimensions in the
exponents; from \cite{Fodor:2012uu}. As shown the obtained values are not compatible.}
\label{conf}
\end{figure}

The conformal fit using the same $\gamma$ in the three channels gave even poorer quality fits than previously in
\cite{Fodor:2011tw} even with the smaller fitted quark mass range. One might attempt to fit each 
channel separately using separate $\gamma$ anomalous dimensions in the exponent. This results in
incompatible values for $\gamma$, see for instance figure \ref{conf} for the tension between the resulting $\gamma$'s from the
pion mass and decay constant channels.

Summarizing the results from the meson spectrum one observes that regardless if one uses 1-loop chiral perturbation theory or polynomial fits a
non-vanishing decay constant and chiral condensate is found in the chiral limit with acceptable fit qualities. 
The power-like conformal fits on the other hand are of poor quality or give non-compatible anomalous dimensions if they are not constrained to be the same.
Hence the analysis in \cite{Fodor:2011tw} and \cite{Fodor:2012uu} favor spontaneous chiral symmetry breaking over conformal infra
red behavior.

There are several ways to improve on the simulations. At weaker coupling $\beta = 3.25$ simulations are ongoing and taste
splitting is much improved over $\beta = 3.2$. The conformal analysis could be improved by considering subleading terms in the
quark mass dependence of the observables. As mentioned already finite size scaling could be used combining information from
several volumes and looking for universal scaling. 

It is worth noting that the small deviation from the leading order behavior $m_\pi^2(m) \sim m$ hints that even if the model is
conformal the anomalous dimension is large. From the scaling $m_\pi(m) \sim m^{1/(1+\gamma)}$ one gets $\gamma \sim 1$. The result
from \cite{DeGrand:2012yq} on the smallness of $\gamma$ is in contradiction with the above observation and it remains to be seen how this
disagreement gets resolved once the necessary improvements are performed to both analyses.

Though the mass spectrum analysis clearly suggest a chirally broken phase, it is important to reduce taste splitting and to
perform a continuum extrapolation by using finer lattices.

\section{Summary and outlook}
\label{summary}

The past several years have seen renewed interest in strong dynamics in the context of the Standard Model. Part of the reason for
it is the realization that higher dimensional representations for the techni fermions may solve some of the problems associated
with constructions based on the fundamental representation. One of the simplest and most appealing constructions is $SU(3)$ gauge
theory with $N_f = 2$ fermions in the sextet representation.

The most basic properties required in order to have any chance for this model to be relevant for phenomenology are all
non-perturbative. Lattice simulations are an ideal tool to address these issues and several methods have been used by several
groups. So far there is no final result which has been carefully extrapolated to the continuum and passed all necessary
consistency checks. The three approaches based on the Schroedinger functional, thermodynamics and mass spectrum disagree in that
the thermodynamics and mass spectrum study seems to indicate the model is chirally broken whereas the latest incarnation of the Schroedinger functional
approach seems to suggest it is conformal. 

The mass anomalous dimension also presents a disagreement: the Schroedinger functional analysis \cite{DeGrand:2012yq} favors a small value $\gamma < 0.45$ whereas
the mass spectrum leads to a much larger $1~\lsim~\gamma~\lsim~2$. Since 
universality is only expected in the continuum it is entirely possible that the disagreements so far are a result of finite
cut-off effects and a more systematic approach of the continuum is necessary.

In this review only the $N_f = 2$ model was mentioned but there are interesting recent quenched results \cite{Fodor:2009ar, Fodor:2009nh}
and also a thermodynamics analysis of the $N_f = 3$ case \cite{Kogut:2011bd}. Both of these cases are useful to study because
unlike in the $N_f = 2$ model it is known what the result should be. The $N_f = 0$ model is chirally broken whereas the $N_f = 3$
model is almost certainly conformal in the infrared. Any method used should reproduce these two features.

\section*{Acknowledgment}

This work was supported by the EU Framework Programme 7 grant (FP7/2007-2013)/ERC No 208740.
I would like to thank Zoltan Fodor, Kieran Holland, Julius Kuti, Chris Schroeder and Ricky Wong for collaboration in the past few years on the
topics covered in this review. I also thank Don Sinclair for providing data from their simulations in order to check the validity
of our code.


\begin{thebibliography}{99}

\bibitem{Sannino:2004qp} 
F.~Sannino and K.~Tuominen,
Phys.\ Rev.\ D {\bf 71}, 051901 (2005)
[hep-ph/0405209].

\bibitem{Dietrich:2006cm} 
D.~D.~Dietrich and F.~Sannino,
Phys.\ Rev.\ D {\bf 75}, 085018 (2007)
[hep-ph/0611341].

\bibitem{Gross:1973id} 
D.~J.~Gross and F.~Wilczek,
Phys.\ Rev.\ Lett.\  {\bf 30}, 1343 (1973).

\bibitem{Politzer:1973fx} 
H.~D.~Politzer,
Phys.\ Rev.\ Lett.\  {\bf 30}, 1346 (1973).

\bibitem{Caswell:1974gg} 
W.~E.~Caswell,
Phys.\ Rev.\ Lett.\  {\bf 33}, 244 (1974).

\bibitem{Jones:1974mm} 
D.~R.~T.~Jones,
Nucl.\ Phys.\ B {\bf 75}, 531 (1974).

\bibitem{Banks:1981nn} 
T.~Banks and A.~Zaks,
Nucl.\ Phys.\ B {\bf 196}, 189 (1982).

\bibitem{Tarasov:1980au} 
O.~V.~Tarasov, A.~A.~Vladimirov and A.~Y.~.Zharkov,
Phys.\ Lett.\ B {\bf 93}, 429 (1980).

\bibitem{Larin:1993tp}
S.~A.~Larin and J.~A.~M.~Vermaseren,
Phys.\ Lett.\ B {\bf 303} (1993) 334
[hep-ph/9302208].

\bibitem{Pica:2010xq} 
C.~Pica and F.~Sannino,
Phys.\ Rev.\ D {\bf 83}, 035013 (2011)
[arXiv:1011.5917 [hep-ph]].

\bibitem{Ryttov:2010iz} 
T.~A.~Ryttov and R.~Shrock,
Phys.\ Rev.\ D {\bf 83}, 056011 (2011)
[arXiv:1011.4542 [hep-ph]].

\bibitem{Peskin:1990zt} 
M.~E.~Peskin and T.~Takeuchi,
Phys.\ Rev.\ Lett.\  {\bf 65}, 964 (1990).

\bibitem{Appelquist:1986an} 
T.~W.~Appelquist, D.~Karabali and L.~C.~R.~Wijewardhana,
Phys.\ Rev.\ Lett.\  {\bf 57}, 957 (1986).

\bibitem{Appelquist:1986tr} 
T.~Appelquist and L.~C.~R.~Wijewardhana,
Phys.\ Rev.\ D {\bf 35}, 774 (1987).

\bibitem{Appelquist:1987fc} 
T.~Appelquist and L.~C.~R.~Wijewardhana,
Phys.\ Rev.\ D {\bf 36}, 568 (1987).

\bibitem{Luscher:1992an} 
M.~Luscher, R.~Narayanan, P.~Weisz and U.~Wolff,
Nucl.\ Phys.\ B {\bf 384}, 168 (1992)
[hep-lat/9207009].

\bibitem{Shamir:2008pb} 
Y.~Shamir, B.~Svetitsky and T.~DeGrand,
Phys.\ Rev.\ D {\bf 78}, 031502 (2008)
[arXiv:0803.1707 [hep-lat]].

\bibitem{DeGrand:2010na} 
T.~DeGrand, Y.~Shamir and B.~Svetitsky,
Phys.\ Rev.\ D {\bf 82}, 054503 (2010)
[arXiv:1006.0707 [hep-lat]].

\bibitem{DeGrand:2012yq} 
T.~DeGrand, Y.~Shamir and B.~Svetitsky,
arXiv:1201.0935 [hep-lat].

\bibitem{Kogut:2010cz} 
J.~B.~Kogut and D.~K.~Sinclair,
Phys.\ Rev.\ D {\bf 81}, 114507 (2010)
[arXiv:1002.2988 [hep-lat]].

\bibitem{Kogut:2011ty} 
J.~B.~Kogut and D.~K.~Sinclair,
Phys.\ Rev.\ D {\bf 84}, 074504 (2011)
[arXiv:1105.3749 [hep-lat]].

\bibitem{Sinclair:2011ie} 
D.~K.~Sinclair and J.~B.~Kogut,
arXiv:1111.2319 [hep-lat].

\bibitem{Gasser:1983yg} 
J.~Gasser and H.~Leutwyler,
Annals Phys.\  {\bf 158}, 142 (1984).

\bibitem{DelDebbio:2010ze} 
L.~Del Debbio and R.~Zwicky,
Phys.\ Rev.\ D {\bf 82}, 014502 (2010)
[arXiv:1005.2371 [hep-ph]].

\bibitem{Fodor:2011tw} 
Z.~Fodor, K.~Holland, J.~Kuti, D.~Nogradi and C.~Schroeder,
arXiv:1103.5998 [hep-lat].

\bibitem{Fodor:2012uu} 
Z.~Fodor, K.~Holland, J.~Kuti, D.~Nogradi, C.~Schroeder and C.~H.~Wong,
arXiv:1205.1878 [hep-lat].

\bibitem{Fodor:2009ar} 
Z.~Fodor, K.~Holland, J.~Kuti, D.~Nogradi and C.~Schroeder,
JHEP {\bf 0911}, 103 (2009)
[arXiv:0908.2466 [hep-lat]].

\bibitem{Fodor:2009nh} 
Z.~Fodor, K.~Holland, J.~Kuti, D.~Nogradi and C.~Schroeder,
JHEP {\bf 0908}, 084 (2009)
[arXiv:0905.3586 [hep-lat]].

\bibitem{Kogut:2011bd} 
J.~B.~Kogut and D.~K.~Sinclair,
Phys.\ Rev.\ D {\bf 85}, 054505 (2012)
[arXiv:1111.3353 [hep-lat]].

\end{thebibliography}
\end{document}